\documentclass[twocolumn,aps,prb,10pt]{revtex4-2}

\usepackage{graphicx}
\usepackage[dvipsnames]{xcolor}
\usepackage{bbm}
\definecolor{bdiv1-00}{HTML}{fcdc64}
\definecolor{bdiv1-01}{HTML}{f8955f}
\definecolor{bdiv1-02}{HTML}{d3583f}
\definecolor{bdiv1-03}{HTML}{9f2720}
\definecolor{bdiv1-04}{HTML}{541511}
\definecolor{bdiv1-05}{HTML}{020102}
\definecolor{bdiv1-06}{HTML}{332346}
\definecolor{bdiv1-07}{HTML}{634487}
\definecolor{bdiv1-08}{HTML}{7375be}
\definecolor{bdiv1-09}{HTML}{7bacef}
\definecolor{bdiv1-10}{HTML}{77edca}

\usepackage[
colorlinks=true,
linkcolor=bdiv1-07,
citecolor=bdiv1-07,
urlcolor=bdiv1-07,
]{hyperref}

\usepackage{stix}
\usepackage{xspace}
\usepackage{amsmath}
\usepackage{amsfonts}
\usepackage{amssymb}
\usepackage{mathtools}
\usepackage{braket}
\usepackage{comment}
\usepackage[capitalize]{cleveref}
\usepackage{sidecap}
\usepackage{chemformula}
\usepackage{enumitem}

\usepackage{makerobust}
\newcommand{\makeauthor}[2]{\newcommand{#1}[1]{{%
  \protect%
  \color{#2}{%
    \bfseries%
    \begingroup\escapechar=-1\edef\x{\endgroup\string#1}\x:%
  }\itshape{} ##1}}%
  \MakeRobustCommand#1}

\newcommand{\supplement}[1]{%
  \clearpage%
  \title{#1}%
  \maketitle%
  \setcounter{equation}{0}%
  \setcounter{figure}{0}%
  \setcounter{table}{0}%
  \setcounter{page}{1}%
  \makeatletter%
  \renewcommand{\thesection}{S\arabic{section}}%
  \renewcommand{\thesubsection}{\Alph{subsection}}%
  \renewcommand{\theequation}{S\arabic{equation}}%
  \renewcommand{\thefigure}{S\arabic{figure}}%
  \renewcommand{\thetable}{S\Roman{table}}%
  \renewcommand{\thepage}{S\arabic{page}}%
  \numberwithin{figure}{section}%
  \numberwithin{table}{section}%
  \numberwithin{equation}{section}%
  \makeatother%
  \onecolumngrid%
}
\makeatletter
\def\maketitle{
\@author@finish
\title@column\titleblock@produce
\suppressfloats[t]}
\makeatother

\let\oldcite\cite
\renewcommand{\cite}[1]{\if\relax\detokenize{#1}\relax\textbf{\color{red}[?]}\else\oldcite{#1}\fi}

\newcommand{\vdagger}{\vphantom{\dagger}}
\newcommand{\bv}[1]{\boldsymbol{#1}}

\makeauthor{\lk}{bdiv1-02}
\makeauthor{\sm}{bdiv1-03}
\makeauthor{\ab}{OliveGreen}
\makeauthor{\mk}{Purple}
\makeauthor{\ji}{blue}
\makeauthor{\md}{cyan}
\makeauthor{\rt}{olive}
\newcommand{\added}[1]{{\color{red}#1}}
\renewcommand{\added}[1]{#1}

\def\paperAuthors{%
    \author{C.~Alexander~Baum}
    \email{christian.baum@uni-wuerzburg.de}
    \author{Jonas~Issing}
    \author{Sarbajit~Mazumdar}
    \author{Matteo~Dürrnagel}
    \author{Michael~Klett}
    \author{Ronny~Thomale}
    \author{Lennart~Klebl}
    \email{lennart.klebl@uni-wuerzburg.de}
    \affiliation{Institut für Theoretische Physik und Astrophysik and Würzburg-Dresden Cluster of Excellence ctd.qmat, Universität Würzburg, 97074 Würzburg, Germany}
    }
\def\paperTitle{%
  Exotic Electronic Order in a Parabolic Kagome Semimetal
}

\begin{document}

\title{\paperTitle}
\paperAuthors

\date{\today}

\begin{abstract}
We study an interacting kagome-lattice realization of a quadratic band-touching semimetal at 2/3 filling with onsite and nearest-neighbor repulsive interactions. Combining functional renormalization group and slave-boson approaches, we map its phase diagram from intermediate to strong coupling and uncover a hierarchy of unconventional electronic orders. The leading instabilities comprise loop-current order, spontaneous altermagnetism arising from a spin-Pomeranchuk instability, and spin-loop-current order with distinctive and largely unexplored properties. We demonstrate how the interplay of band kinematics, electronic interactions, and quantum geometry governs the selection of these phases. Our findings establish quadratic band-touching semimetals as a promising platform for unconventional symmetry breaking and suggest analogous phenomena in other parabolic semimetals.
\end{abstract}

\maketitle

\section{Introduction}
Parabolic bands, i.e., areas of band momentum where the effective mass is constant as a function of lattice momentum $\bv k$, are the characteristic signature of gapped semiconductors around their valence band maximum and conduction band minimum, and canonically suggest a continuum description of effectively free electronic quasiparticles at low energies. The scenario becomes substantially different as one considers a zero gap semiconductor with quadratic band touching~\cite{PhysRev.102.1030, abrikosov1971possible, abrikosov1974calculation} which we frame as parabolic semimetals in the following: placing the Fermi level at the quadratic band touching, the large density of states produces an enhanced propensity for electronic order as a mechanism to remove electronic spectral weight from the Fermi level. Instances for parabolic semimetals have been proposed in one spatial dimension such as for polyacene~\cite{PhysRevB.28.7236,PhysRevB.88.224512,Schmitteckert2017i} which is tuned such that Dirac cones of opposite chirality annihilate. The parabolic band touching in such systems, with its diverging density of states $\mathrm{DOS}(E)\propto |E-E_F|^{-1/2}$, possesses stability against some perturbations such as higher order dispersion variations that leave the chiral symmetry intact. Parabolic semimetals in two spatial dimensions~\cite{sun2009topological, dora2014occurrence} have been prominently discussed in \added{twisted~\cite{ingham2023quadraticdirac} and} Bernal bilayer graphene~\cite{PhysRevLett.96.086805, vafek2010many}. Here the topological winding number adds to the stability of the quadratic band touching against most perturbations that preserve chiral symmetry, except for trigonal warping.

Electronic states on the kagome lattice have attracted tremendous interest over the past decade~\cite{Neupert2022c,DiSante2026}. Long regarded primarily as a paradigm of frustrated magnetism, and often encountered as the two-dimensional (111) plane of the pyrochlore lattice, kagome systems were traditionally studied in the context of strongly correlated quantum magnets~\cite{RevModPhys.88.041002}. The theoretical prediction~\cite{Kiesel2012s,Kiesel2013u} and subsequent experimental discovery~\cite{PhysRevMaterials.3.094407,PhysRevLett.125.247002} of kagome metals have, however, established these materials, and other kagome compounds in general, as a broad and versatile platform for exploring the interplay of topology, electronic correlations, and unconventional ordered phases in contemporary condensed matter physics. Why this is so already becomes suggestive from its single-particle band structure.
Depending on the Fermi level tuning by chemical composition, kagome materials exhibit band features as diverse as flat bands, Dirac cones, and Van Hove singularities. From there, a plethora of intriguing yet previously largely elusive electronic phenomena have been theoretically proposed and/or experimentally discovered in kagome metals, including exotic loop current orders, superconductivity, strongly correlated Dirac fluids, room temperature anomalous Hall effect, and many more~\cite{ye-n2018,PhysRevLett.125.247002,mielke-n2022,mazin-nc2014}. To date, most of these widely discussed phenomena favor a Fermi level positioned at the Dirac point or near Van Hove singularities.

In this article, we argue that yet another realm of highly exotic electronic order is to be found at flat band level on the kagome lattice. The kagome flat band is of Mielke-Tasaki type~\cite{mielke1993}, i.e., it emerges due to destructive interference of individual electronic states leading up to localized electronic states that become a part of the flat band. It is important to note that the Mielke construction for a droplet of $N$ kagome unit cells produces $N-1$ localized states, and imposing toroidal boundary conditions adds another two non-contractible loop states~\cite{Bergman2009b,PhysRevB.93.155155,Rhim2019c}, totalling $N+1$ degenerate states.  This accounts for the parabolic band touching between the flat kagome band and the adjacent dispersive band. By introducing some dispersivity to the flat band, we manage to recover a parabolic semimetallic setting on the kagome lattice with a stability mechanism inherited from the flat band, rather than from Dirac cone annihilation as in polyacene or topological winding in Bernal bilayer graphene. As we investigate the ordering instabilities of this parabolic kagome semimetal exposed to onsite and nearest neighbor repulsive interactions, we reveal a whole series of exotic electronic orders such as charge loop-current order, spin Pomeranchuk instabilities leading up to spontaneous altermagnetic order, and spin loop current order. We find that parabolicity, the quantum distance of Fermi level eigenstates due to sublattice interference~\cite{Kiesel2012s}, and interactions in the parabolic kagome semimetal conspire to reveal largely unprecedented electronic order which promises to open a new chapter of electronic ordering phenomena in kagome materials.

\section{Model}
\begin{figure}
    \centering
    \includegraphics[width=\linewidth]{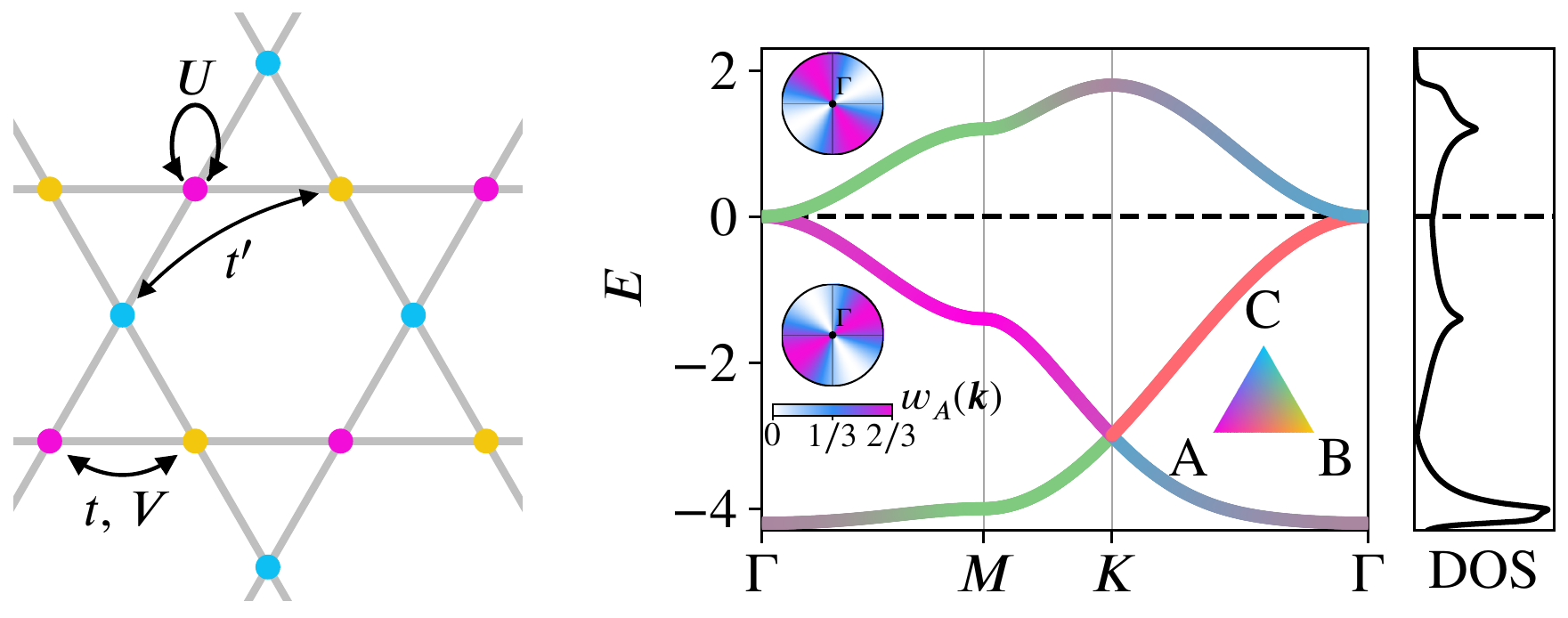}
    \caption{Real-space picture of the generalized kagome Hubbard model (left) and corresponding non-interacting band structure (right).
    We indicate the nearest-neighbor hopping $t$, next-nearest-neighbor hopping $t'$, onsite Hubbard interaction $U$, and nearest-neighbor density-density interaction $V$ (left panel). The right panel shows the corresponding band structure for $t=1$, $t'=-0.3$ at filling $\nu=2/3$, with the color encoding the sublattice character (A: magenta, B: yellow, C: blue) of the Bloch states. The insets show the Bloch state weight of sublattice A ($w_A(\bv k) = |u_A(\bv k)|^2$) in the vicinity of $\bv k=\Gamma$ for the topmost (upper inset) and middle (lower inset) bands. We note that the bands' characters do \emph{not} become fully pure, instead they smoothly interpolate between the two possible $E_2$ representations upon encircling $\bv k=\Gamma$.
    We also plot the non-interacting density of states (DOS) on the right, highlighting that at $E=0$ the parabolic bands induce a nonzero DOS.
    }
    \label{fig:model}
\end{figure}
We study the generalized kagome Hubbard model (KHM)
\begin{equation}
\begin{multlined}
    \mathcal{H} =
    -\sum_{ij\sigma} \big(t \delta_{\langle i,j\rangle} + t' \delta_{\langle\!\langle i,j\rangle\!\rangle} \big)
    \big( c_{i\sigma}^\dagger c_{j\sigma}^{\vdagger} + \mathrm{h.c.} \big)\\
    + U\sum_{i}n_{i\uparrow}n_{i\downarrow}
    + \frac{V}{2}\sum_{\langle i,j\rangle\sigma\sigma'}
    n_{i\sigma}n_{j\sigma'},
\end{multlined}
\label{eq:Model}
\end{equation}
where $c^\dagger_{i\sigma}, c_{i\sigma}^{\vdagger}$ denote electron creation and annihilation operators at site $i$ and spin $\sigma$, $n_{i\sigma} = c^\dagger_{i\sigma} c_{i\sigma}^{\vdagger}$, $t$ ($t'$) and $U$/$V$ the (second) nearest-neighbor (NN) hopping as well as on-site/nearest-neighbor interaction. We define energy units by setting the nearest-neighbor hopping to $t=1$. The kagome lattice geometry is depicted in \cref{fig:model} with its three sublattices colored in magenta (A), yellow (B), and blue (C).
At $t'=0$, it features two Van Hove singularities at the $M$ point with pure, single-orbital ($p$-type) and mixed-orbital ($m$-type) character as well as a flat band that is made dispersive by the inclusion of $t'\neq 0$. To tune the system to a semimetallic regime, we set the filling to $\nu=2/3$ and $t'<0$---reminiscent of realistic band structures for, e.g., chromium-based kagome metals~\cite{liu2024superconductivity, Liu2026, Crispino2025t, Chatzieleftheriou2026p}.

The $\nu=2/3$ filling fixes the Fermi level to the parabolic band touching point at $\bv k=\Gamma$, see dashed line in band structure in \cref{fig:model}. The upper (particle-like) and lower (hole-like) band can be made particle-hole symmetric in the parabolic regime at $t'=-1/3$, hinting towards enhanced particle-hole fluctuations. Even though the Fermi surface collapses to a single point (at $\bv k=\Gamma$), the system hosts rich structure encoded in its quantum geometry. For the kagome Hubbard model tuned to the Van-Hove points, the interplay of quantum geometry (or orbital makeup) and energetics (i.e., Fermi surface, density of states) results in the ``sublattice interference''{} mechanism for various unconventional orders mediated by Hubbard $U$ and nearest-neighbor $V$~\cite{Kiesel2012s}. Here, in contrast, we tune the system to the parabolic band touching point at $\bv k=\Gamma$, which effectively masks out effects stemming from momentum space structure.

This leaves us with an ideal platform to study electron-electron interactions in a parabolic semimetal with nontrivial Bloch states (that together span the $E_2$ irreducible representation of the $C_{6v}$ point group). We highlight that the interplay of quantum geometry and interactions is present in a pure form
by lack of an extended Fermi surface. In the following, we uncover various unconventional charge- and spin instabilities in the model by a complementary weak-to-intermediate coupling treatment with the Functional Renormalization Group (FRG, see \cref{ssec:frg}) and a strong-coupling perspective through slave-boson mean-fields (SBMF, see \cref{ssec:sbmf}).

\section{Methods}
FRG~\cite{Metzner2012, Platt2013, Dupuis2021} is suitable for tracking phase transitions of itinerant electrons in the weak-to-intermediate coupling regime without mean-field bias. As it cannot trivially be extended to the strong coupling regime, we complement it with slave-boson mean-field (SBMF) techniques to also access the strong coupling regime that naturally emerges in parts of the FRG phase diagram (cf.~\cref{fig:Phase_diagram}). This section gives a brief introduction to both methods, which are then applied to the KHM in \cref{sec:results}. 

\subsection{Functional Renormalization Group}
\label{ssec:frg}
In FRG, the evolution of the two-particle vertex from ultraviolet to infrared cutoff is expressed via a continuous flow equation.
The scattering contributions of the vertex are decomposed into the three diagrammatic channels~\cite{Lichtenstein2017}:
\begin{equation}
\begin{multlined}
     V^\Lambda(\bv{k}_1,\bv{k}_2,\bv{k}_3,\bv{k}_4)
     =
     V_0
     +
     P^\Lambda(\bv k_1 + \bv k_2; \bv k_1, \bv k_3)\\
     +
     C^\Lambda(\bv k_1 - \bv k_4; \bv k_1, \bv k_3)
     +
     D^\Lambda(\bv k_1 - \bv k_3; \bv k_1, \bv k_4),
\end{multlined}
\label{eq:VertexDecomposition}
\end{equation}
where $\Lambda$ corresponds to the renormalization group scale, and $P$, $C$, and $D$ label particle-particle, crossed and direct particle-hole channels, respectively. The first momentum dependency is the transfer momentum native to the channel, i.e., a bosonic momentum expressed as sum/difference of two fermionic momenta.
In the $SU(2)$ symmetric (spin rotationally invariant) case, the FRG flow equation written in terms of the three diagrammatic channels reads
\begin{align}
    \label{eq:frg-compact}
    \dot V^\Lambda &{}= \dot P^\Lambda + \dot C^\Lambda + \dot D^\Lambda \,, \\
    \label{eq:frg-compact-P}
    \dot P^\Lambda &{}= V^{\Lambda, P} \circ \dot \chi^{\Lambda, P} \circ V^{\Lambda, P} \,, \\
    \label{eq:frg-compact-C}
    \dot C^\Lambda &{}= V^{\Lambda, C} \circ \dot \chi^{\Lambda, H} \circ V^{\Lambda, C} \,, \\
    \label{eq:frg-compact-D}
    \dot D^\Lambda &{}= \begin{multlined}[t]
        -2\left( V^{\Lambda, D} - \frac12V^{\Lambda,C} \right) \circ \dot \chi^{\Lambda, H} \\ {}\circ \left( V^{\Lambda, D} - \frac12V^{\Lambda,C} \right) + \frac12 \dot C^\Lambda \,,
    \end{multlined}
\end{align}
where ``$\circ$''{} resembles a matrix product over all quantum numbers, $V^{\Lambda,X}$ denotes the vertex projected to channel $X \in \{P,C,D\}$ and $\dot \chi^{\Lambda, P/H}$ is the particle-particle/particle-hole loop derivative:
\begin{equation}
\begin{aligned}
    \dot\chi^{\Lambda, P/H}(q,k) &{}= \dot G^\Lambda(k) G^\Lambda(\pm q\mp k) + \dot G^\Lambda(\pm q\mp k) G^\Lambda(k) \\
    &{}= \begin{multlined}[t]
    \frac12 \Big[ \delta(\Lambda - |ik_0|)
    \, G(k) G(\pm q\mp k) \\
    {} + \delta(\Lambda - |iq_0-ik_0|)\, G(\pm q\mp k) G(k) \Big] \,.
    \end{multlined}
\end{aligned}\label{eq:loopdev}
\end{equation}
Note that the form of $\dot{\chi}$ in \cref{eq:loopdev} already encodes the choice of sharp frequency cutoff as regulator~\cite{Beyer2022r}, naturally making the RG scale $\Lambda$ an energy scale. We treat the FRG flow equation \cref{eq:frg-compact} in the common approximations of (i)~discarding frequency dependencies of the two-particle vertices and (ii)~neglecting the flow of the one-particle vertex (self-energy), suitable for instability analysis~\cite{salmhofer2001fermionic,Metzner2012}.

The flow equations of the three diagrammatic channels \cref{eq:frg-compact-P,eq:frg-compact-C,eq:frg-compact-D} extend a single-channel, random phase approximation (RPA) approach to an unbiased multi-channel analysis where cross-channel effects are incorporated---overcoming the Fierz ambiguity of selecting a distinct channel to bosonize.
To express the vertex functions efficiently, we employ the truncated-unity (TU) approximation known as TUFRG~\cite{Husemann2009e, Lichtenstein2017, Beyer2022r, Profe2022t} and make use of its implementation in the open-source library {divERGe}~\cite{10.21468/SciPostPhysCodeb.26, 10.21468/SciPostPhysCodeb.26-r0.5}.
There, the fermionic momentum dependencies of each of the channels [cf.~\cref{eq:VertexDecomposition}] are written in a truncated real-space form factor basis, while the bosonic transfer momentum is kept explicitly. Physically, the TU approximation corresponds to assuming that fluctuations between electronic bilinears are constrained to short-range distances.

As the energy scale $\Lambda$ is lowered, the vertex picks up scattering corrections that drive the Fermi liquid towards symmetry-breaking order. Eventually, the flow of $V^\Lambda$ breaks down due to a divergence of vertex elements. Performing an eigendecomposition of the leading channel close to the divergence corresponds to solving the linearized gap equation at the transition---with an appropriately renormalized vertex that includes cross-channel effects. The unbiased nature of the FRG flow therefore allows us to distinguish between pairing, magnetic and charge order parameters with arbitrary orbital, spin, and real space (bond) structure. Moreover, it allows us to track the evolution of vertex elements as the scale is lowered, which enables us to obtain insight into the driving processes for a given order as well as competition between different orders.

\subsection{Slave-boson mean-field}
\label{ssec:sbmf}
\begin{figure*}
    \centering
    \includegraphics{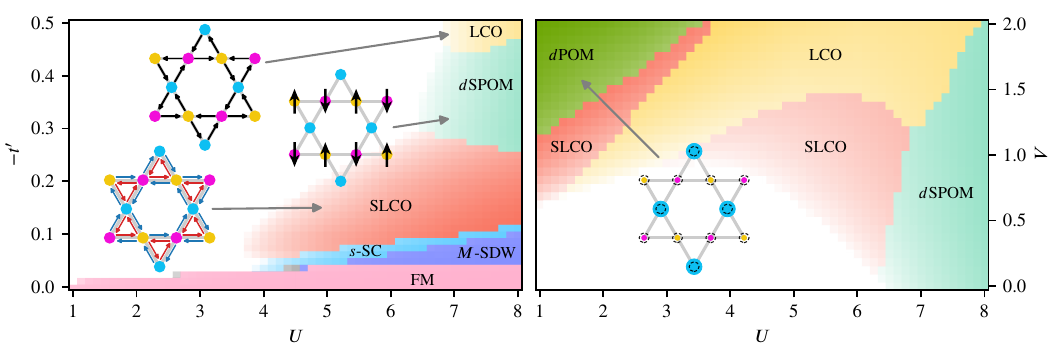}
    \caption{
    FRG phase diagram for varied $t'$, $U$ (at $V=0$, left) and $U$, $V$ (at $t'=-0.3$, right).
    The color hue encodes the phase, while the intensity corresponds to the critical scale (white: no order, i.e., parabolic semimetallic Fermi liquid phase).
    The left panel includes visualizations of the loop current order state (LCO, yellow), the $d$-wave spin Pomeranchuk instability ($d$SPOM, aqua), and the spin loop current order state (SLCO, red), where red and blue arrows correspond to the spin-$\uparrow$ and spin-$\downarrow$ loop current, respectively. The lower part of the phase diagram, where the topmost band becomes flat is occupied by a flat-band-ferromagnet (FM, pink), which is flanked by a $\bv q=M$ spin density wave ($M$-SDW, dark blue), an extended $s$-wave superconductor ($s$-SC, light blue), and tiny regions of $\bv q=K$ charge density wave (gray)---note that FRG becomes unreliable in the $t'\to0$ limit.
    Introducing nearest neighbor repulsion $V$ (right panel) triggers a $\bv q = 0$ charge density wave instability ($d$POM, dark green; $d$-wave Pomeranchuk), which features inequivalent population of sublattices. For the $d$POM and $d$SPOM states, the insets display one representative of the two-dimensional $E_2$ irreducible representation, which FRG does not single out as symmetries are kept intact. We note that the shown representatives are those who minimize the free energy in slave boson mean-field theory (cf.~\cref{sec:strong-coupling}).
    }
    \label{fig:Phase_diagram}
\end{figure*}

To access the strongly correlated regime beyond the weak-coupling description, we employ the spin-rotationally invariant Kotliar-Ruckenstein (KR) slave-boson (SB) formalism \cite{Kotliar1986n, Fresard1997i, Lechermann2007r}. Within this method the electronic operator is written as 
\begin{equation}
    c^\dagger_{i\sigma}
    =
    \sum_{\varsigma}
    z^{\dagger}_{i,\sigma\varsigma}
    f^\dagger_{i\varsigma},
\end{equation} 
where \(f_{i\varsigma}\) denotes the pseudofermion and  $ z_{i}$  is the KR quasiparticle-renormalization matrix. This representation is obtained by enlarging the local Hilbert space with auxiliary bosonic fields describing empty, singly and doubly occupied configurations---and the pseudofermions $f^{(\dagger)}$, which must be introduced to keep the electronic operators' fermionic character. The physical subspace is then recovered by imposing local constraints on the bosonic fields through Lagrange multipliers. To describe the kagome lattice with its multi-sublattice unit cell and to allow for sublattice-dependent charge and magnetic configurations, we employ a cluster mean-field ansatz, where the ground state of the system is then given as the correlated saddle point of the free energy $F$ with respect to the bosonic fields and Lagrange multipliers. 

In addition to the ground state solution, we consider static Gaussian fluctuations around the correlated saddle point. The corresponding fluctuation matrix yields sublattice-resolved spin and charge susceptibilities, which we use to assess possible instabilities of the SBMF saddle point ~\cite{Riegler2020s,Seufert2021b, Riegler2023i,Issing2026a}.

\section{Results}
\label{sec:results}
We perform FRG simulations of the $U$--$t'$ and $U$--$V$ phase diagrams of the KHM at $\nu=2/3$ filling (see \cref{app:frg-details} for parameters) and corroborate them in the strongly coupled regions (i.e., large $U$ or $V$) with SBMF theory (see \cref{app:sbmf-details} for parameters).

\subsection{FRG analysis}
\Cref{fig:Phase_diagram} shows the leading instabilities of the KHM at $\nu = 2/3$ as a function of onsite repulsion $U$ and second neighbor hopping $t'$ as obtained from FRG.
At small $t'$, intermediate coupling effects gain importance due to the (almost) flat band at the Fermi level.
In this regime, FRG predicts---in addition to the flat-band ferromagnet (FM) at $-t' \gtrsim 0$---an $M$-point modulated spin density wave ($M$-SDW) with admixtures of a spin bond order (SBO).
This state is induced by ``diffuse nesting''~\cite{Johannes2008f,Beck2025k} at $\bv q=M$ stemming from the $m$-type Van Hove singularity in the topmost band, which is relevant at small but finite $t'$. It is compatible with the well known SDW+SBO phase of the KHM at $m$-type Van Hove filling~\cite{Profe2024k, Wenger2025t}. Upon increasing the energetic distance of the $m$-type Van Hove singularity to the Fermi level, the diffuse nesting is weakened and the $M$-SDW gives way to an extended $s$-wave superconductor ($s$-SC) with dominant weights on the third neighbor bonds. \added{By $E_2$ symmetry of the involved Bloch states, we expect a superconductor that gaps the Fermi surface to lie in one of the following irreducible representations (irreps): $A_1 \oplus A_2 \oplus E_2 = E_2\otimes E_2$. The onsite repulsion $U$ is avoided by extending in the bond sector exactly on those bonds connected by a sign change in the $M$-SDW. As a result the $A_2$ state ($i$-wave) is suppressed; it does not have support on the spin fluctuation spectrum (these bonds are symmetry forbidden in the $A_2$ irrep). The $E_2$ state ($d+id$-wave) fails to gap the Fermi surface to linear order in $\Delta$, so the $A_1$ state ($s$-SC) remains as seen in the FRG.}

Since the phases discussed above reside in a regime where $U$ is large compared to the relevant bandwidth of the electronic states participating in the symmetry breaking transition, they are reconsidered using SBMF as stronger-coupling technique in \cref{sec:strong-coupling}.
We hence focus our main attention to the regime of larger $t'$ in the following, where the increased bandwidth renders an FRG treatment more reliable.
In this regime, the Fermiology more strongly determines electronic fluctuations, which can also be seen by the exclusive appearance of $\bv q = 0$ phases away from the $-t'\gtrsim0$ line.
This gives rise to antiferromagnetic (AFM) spin fluctuations (encoded in nearest-neighbor spin exchange $J$) driven by onsite repulsion $U$ as depicted in \cref{fig:diagrams}(a).
When accounting for further renormalization, i.e., many-body effects, these underlying AFM fluctuations drive other exotic phases at $\bv q = 0$: a loop-current order (LCO) that breaks time reversal symmetry $\mathcal T$, and a \emph{spin}-loop-current order (SLCO), which breaks parity $\mathcal P$.

\subsubsection{Spin and Charge loop current orders}

\begin{figure}
    \centering
    \includegraphics[width=\linewidth]{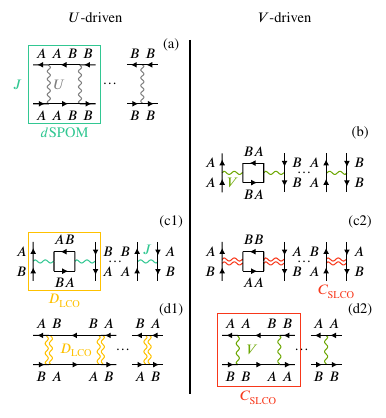}
    \caption{
    Diagrams contributing to the (spin) Pomeranchuk instabilities, and (spin) loop current orders (from top to bottom) in the case of large $U$ and $V$.
    Through cross-channel renormalization, nematic vertex corrections $J$ associated with the $d$SPOM state can drive LCO $\sim J$ and SLCO $\sim J^2$.
    Likewise for $V>0$ nematic charge fluctuations associated with $d$POM components can drive SLCO $\sim V$ as well as LCO  $\sim V^2$.
    We show the $A/B$ sublattice components for clarity but symmetry under permutation of sublattice indices is implied.
    In (d1) and (c2) we show only the (A,B) $\leftrightarrow$ (B,A) vertex components, since these provide the cross-channel feedback necessary for (c1) and (d2) respectively.
    Note that the internal propagators which enter the diagrams are sublattice-mixed and -pure; both supported by the Bloch states' character with similar magnitude (cf.~\cref{fig:model}).
    }
    \label{fig:diagrams}
\end{figure}

In \cref{fig:diagrams} we identify the lowest order diagrammatic contributions promoting the emergence of (S)LCO orders in the FRG.
For Hubbard-$U$ only (cf.~left panel of \cref{fig:Phase_diagram}), both rely on the underlying nearest-neighbor antiferromagnetic $J$, but to different orders:
The LCO is already supported on the 1-loop level when inserting the spin exchange interaction into the charge ($D$) channel in \cref{fig:diagrams}(c1).
This generates an interaction $D_\text{LCO}$ that directly gives a Stoner-type divergence when inserted as central kernel in an RPA-like ladder resummation as indicated in \cref{fig:diagrams}.
For the SLCO, such an attractive channel is only realized after feeding back $D_\text{LCO}$ to the magnetic $C$ channel \cref{fig:diagrams}(d1). This results in an attractive scattering vertex for this order at one loop order higher than for the LCO.

This is also directly evident when analyzing the sublattice structure of the diagrams in the $C$ and $D$ channel, respectively:
The AFM exchange $J$ describes interactions between spin densities on adjacent sites and is native to the $C$ channel, which would translate to a \emph{single} bosonic momentum dependence in the $C$ channel [see \cref{eq:VertexDecomposition}].
Bond-only orders, such as (S)LCO, on the other hand translate to momentum dependencies in the fermionic momenta. These cannot be generated within the same channel but are mediated by projecting the interaction in one of the other diagrammatic channels via \cref{eq:frg-compact-P,eq:frg-compact-D}---to which the interaction is not native.
After an appropriate bond interaction $D_\text{LCO}$ has been generated [cf.~\cref{fig:diagrams}(c1)], backfeeding it to the $C$-channel by means of \cref{fig:diagrams}(d1) can also seed a loop current order in the magnetic channel, i.e., an SLCO.

The hierarchy of loop orders is reflected in the FRG phase diagram of \cref{fig:Phase_diagram}.
The low-lying density of states (DOS) of the upper band scales with the parabolic band curvature $ \propto 1 / |t'|$ and consistently decreases from the divergent flat band limit as $t'$ is increased. 
This directly affects the value of the fermionic loops at $\bv q = 0$ [cf.~\cref{eq:loopdev}] in the diagrammatic expansion of \cref{fig:diagrams} that scale like the DOS at the Fermi level and hence allows to relate the relevant loop orders with the value of $t'$.
At large $t'$---or, equivalently small loops---higher loop orders like \cref{fig:diagrams}(c1) are suppressed and the 1-loop correction promotes the LCO as dominant instability.
As $t'$ is reduced higher order loop corrections eventually overtake the 1-loop contributions to drive the SLCO phase.

The inclusion of NN Coulomb repulsion $V$ opens up new channels for the generation of (S)LCO.
On the bare level, this interaction supports an $l = 2$ Pomeranchuk instability ($d$POM) in the $E_2$ irreducible representation (irrep).
This corresponds to an occupation imbalance of the respective sublattice sites originating from the geometrical frustration of the NN repulsion (see inset in \cref{fig:Phase_diagram}) and represents the generic symmetry breaking propensity of the kagome lattice in the $V$-dominated limit not only at Van Hove~\cite{Kiesel2013u,Wang2013c,Profe2024k,Zhan2026l} but also incommensurate fillings~\cite{Bigi2025p}.

Since $V$ is related to $J$ via fermionic exchange symmetry it features the same structure but is native to the $D$ instead of the $C$ channel. It therefore gives rise to the same higher order corrections as $J$, just with the $C$ and $D$ channels interchanged. As a consequence its support for the different phases is inverted as well:
While $V$ couples to the SLCO at 1-loop order via a single cross-channel projection, an additional loop insertion is required to couple to the LCO, just like for the SLCO in the $J$-only case.
We showcase the $J \leftrightarrow V$ relation in the two columns of diagrammatic expressions depicted in \cref{fig:diagrams}.
As the generic leading-order effect from the Hubbard-$U$ is the 1-loop induced $J$ [cf.~\cref{fig:diagrams}(a)], we can understand the location of phases in the $U$--$V$ phase diagram (cf.~right panel of \cref{fig:Phase_diagram}) from the $J\leftrightarrow V$ relation:
In the $V$-dominated regime, the SLCO is directly supported by the 1-loop diagram in \cref{fig:diagrams}(d2).
Upon increasing $U$ (and the induced $J$), we observe a competition between the $V$- and $J$-driven effects that leads to a suppression of the SLCO scales (color intensity in \cref{fig:Phase_diagram}).
Hence, the competition of two diagrammatic contributions ($U$ and $J$ vs.~$V$) has a similar effect to the reduction of low-lying DOS via sizable $t'$, where the lowest loop orders in the electronic fluctuations take over and promote LCO as the dominant instability.
When entering the regime of large $U/V$, the AFM exchange $J$ presents the only dominant interaction process and higher loop orders drive again an SLCO state.

\begin{figure}
    \centering
    \includegraphics{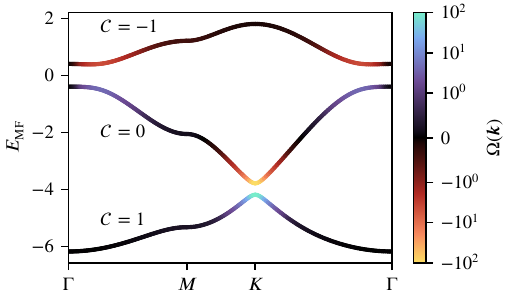}
    \caption{Quasiparticle band structure in the loop current phase (LCO) obtained from inserting the LCO eigenvector obtained from FRG as an (ad hoc) mean-field. The bands are colored corresponding to their Berry curvature $\Omega(\bv k)$ and we indicate their Chern number $\mathcal C$ on the left. Note that only the absolute value and relative sign of $\mathcal C$ is fixed, as the LCO state features a residual $\mathbb Z_2$ symmetry.}
    \label{fig:lco}
\end{figure}

Notably, the observed (S)LCO states do not break the crystalline translation symmetry but break time reversal $\mathcal T$ (parity $\mathcal P$) while preserving the other one.
This difference is solely induced by the presence of finite magnetization on the bonds in the SLCO. The real space structure remains otherwise identical leading to the same (spin) Berry curvature in the quasi-particle band structure of the ordered phase (cf.~\cref{fig:lco}).
Since the current orders gap the band completely, one obtains isolated bands with non-trivial (spin) Chern number, that are topologically distinct from previous reports of LCO induced Chern bands on the kagome lattice at Van Hove filling, where the $\bv q = M$ Fermi surface nesting enables translation symmetry breaking LCO~\cite{Jiang2021u, Wang2021e, Denner2021a, Neupert2022c, Zhan2026l} and SLCO orders~\cite{Wenger2025t}.
\added{
The phenomenology of SLCO and a more detailed account of its potential microscopic origins will be presented in Ref.~\cite{ingham2026spinloop}.
}

\subsubsection{Spin Pomeranchuk instability at particle hole symmetric filling}
In between the (S)LCO orders in the $t'$--$U$ phase diagram (around $t'=-1/3$, left panel of \cref{fig:Phase_diagram}), we find a flow to strong coupling in the $d$SPOM channel. In this region, cross-channel feedback promoting the loop current phases is heavily suppressed. Since the quadratic band touching point at $\bv k=\Gamma$ becomes particle-hole symmetric at $t'=-1/3$, the particle-particle bubble and the particle-hole bubble at $\bv q = 0$ are degenerate, which leads to a cancellation of contributions in the $P$ and $C/D$ channels.
This results in the diagrammatic expansion of the effective vertex being almost exclusively given by the AFM coupling $J$ depicted in \cref{fig:diagrams}(a) generated at 1-loop order in $U$.
Without inter-channel effects, this interaction in the magnetic channel ($C$) takes up an equivalent role to the NN Coulomb repulsion $V$ in the charge channel ($D$) and fosters a $d$-wave spin Pomeranchuk instability ($d$SPOM) in the $E_2$ irrep with equal real space structure to the $d$-wave Pomeranchuk order ($d$POM).
For the latter, the suppression of current orders via $P$ channel feedback has already been observed in the spinless kagome Hubbard model at Van Hove filling~\cite{Zhan2026l}.

\begin{figure}
    \centering
    \includegraphics{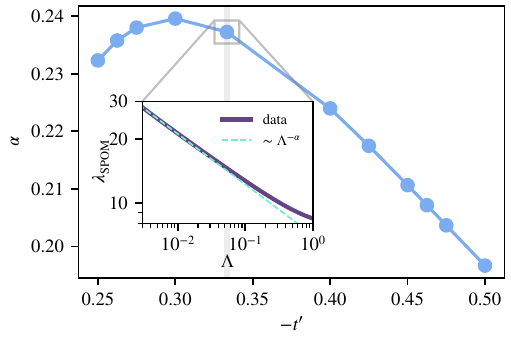}
    \caption{Power-law exponent $\alpha$ of the $d$SPOM fluctuations within FRG as a function of $t'$ for $U=8$. The FRG flow of the $d$SPOM vertex eigenvalue $\lambda_\mathrm{SPOM}$ is fitted with a power law ($\log\lambda_\mathrm{SPOM} \approx -\alpha\log\Lambda + m$) in the region $\Lambda \in [ 3 \cdot 10^{-3} , 5 \cdot 10^{-3} ]$, i.e., at low RG scales. The inset displays one such fit at the approximately particle-hole symmetric point $t'=-1/3$, which roughly coincides with the maximum of $\alpha(t')$.}
    \label{fig:power}
\end{figure}

While this picture suggests a Stoner-like divergence coinciding with a freezing of the dominant nematic spin fluctuations, the FRG flows in the $d$SPOM regime in \cref{fig:power} only feature a power-law divergence of the corresponding eigenvalue of the effective two particle vertex $\lambda_\mathrm{SPOM}\sim \Lambda^{-\alpha}$.
The exponent $\alpha$ is enhanced around the particle-hole symmetric point in \cref{fig:power}, but a weak coupling divergence is only recovered when diagrammatic contributions from the $P$-channel are discarded.
This hints at residual cross-channel suppression of the $d$SPOM via pair fluctuations even in the presence of particle-hole symmetry (which enhances particle-hole fluctuations, i.e., $d$SPOM).

We reinforce the above hypothesis by slightly doping the system to $\mu = \pm 10^{-2}$. The loss of particle-hole symmetry has the effect of breaking the  algebraic divergence at scales $\Lambda \sim |\mu|$, which highlights the importance of inter-band nesting at $\bv q = 0$ due to particle-hole symmetry for the $d$SPOM instability.
This close correspondence yields interesting parallels with excitonic condensates at vanishing transfer momentum: As the particle and hole bands of excitonic condensates typically feature an almost perfect orbital polarization, they allow for a clear definition of the order parameter in terms of inter-orbital charge order~\cite{Jerome1967, Uebelacker2011, Kaneko2025}.
This stands in contrast to the band structure of the KHM at $2/3$ filling in \cref{fig:model}, which shows a strong momentum dependence of the orbital character around $\bv k=\Gamma$ for both the particle and hole band spanning the two $E_2$ eigenstates of $C_{6v}$ that are inherently mixed.
Since this non-trivial quantum geometry also enters the (non-interacting) susceptibility through the orbital-to-band transformation, it suppresses conventional excitonic physics giving way to the $d$SPOM phase.

\begin{figure*}
    \centering
    \includegraphics[width=0.49\linewidth]{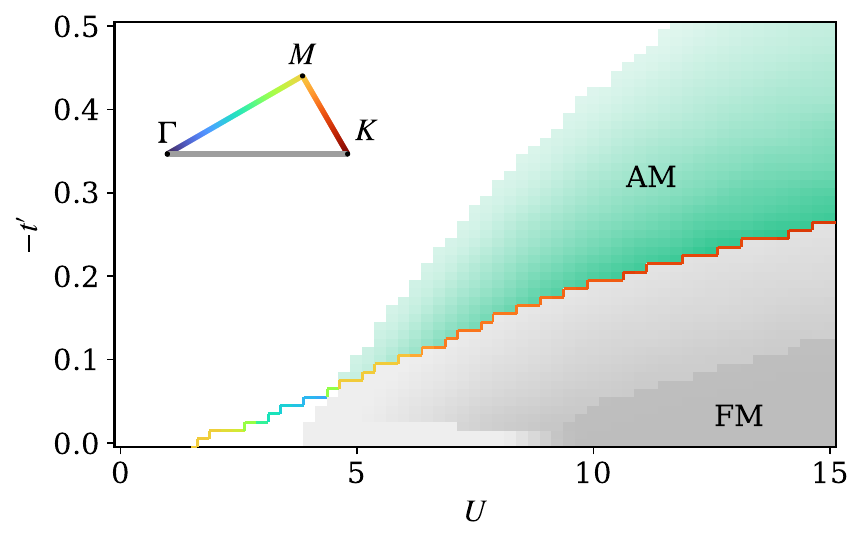}
    \hfill
    \includegraphics[width=0.49\linewidth]{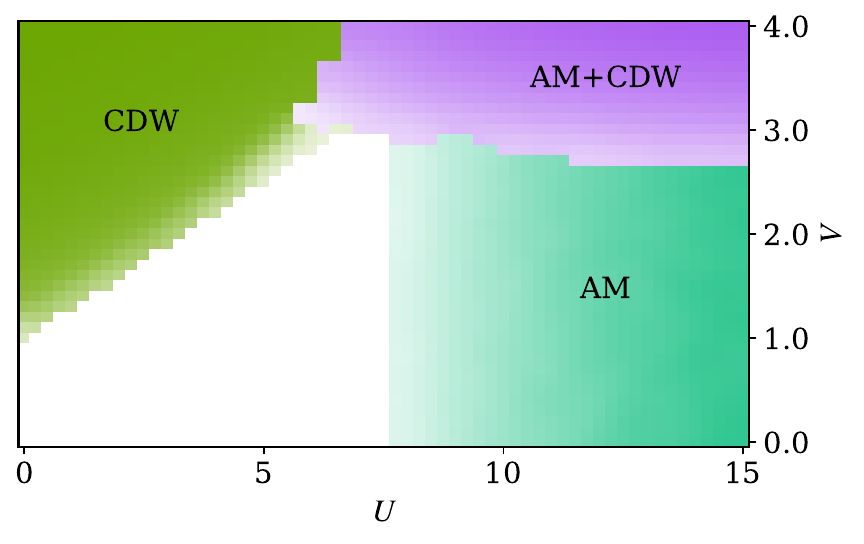}
    \caption{
    SBMF phase diagrams for varied $t'$, $U$ (at $V=0$, left) and $U$, $V$ (at $t'=-0.3$, right). The color hue encodes the phase, while the intensity corresponds to the size of the order parameter. The left panel shows a ferromagnetic (FM) state for small $|t'|$, favored by the nearly flat band, while for larger negative $t'$ the system develops a nematic
    altermagnetic (AM) spin order consistent with the $d$SPOM identified in the FRG analysis. The grey region indicates where the correlated saddle point becomes unstable with respect to transverse spin fluctuations. The colored line encodes the ordering vector $q$ of the leading instability. In the right panel, onsite repulsion predominantly drives the AM phase, while
    nearest-neighbor repulsion favors an intra-unit-cell charge-density-wave
    (CDW) order transforming in the $E_2$ representation. Between these
    limits, spin and charge nematic order coexist, resulting in a combined
    AM+CDW phase.
    From the evolution of the order parameters, we identify the
    PM--AM
    \added{transition}
    as continuous, second-order,
    whereas the transitions into the \added{CDW and} AM+CDW coexistence phases are
    discontinuous and hence first order.
    }
    \label{fig:Phase_diagram_SBMF}
\end{figure*}

Inside the symmetry broken phase, the order parameter lies within the two dimensional $E_2$ irrep spanned by two basis vectors that can be classified by their transformation behavior under vertical and horizontal mirror operations as even or odd.
While all linear combinations within the eigenspace are degenerate directly at the phase transition, this degeneracy is broken when considering a finite order parameter and the system realizes a linear combination that minimizes the free energy most efficiently.

\subsection{The Spin Pomeranchuk instability from strong coupling}
\label{sec:strong-coupling}

The algebraic divergence of the \(d\)SPOM eigenvalue observed in the FRG analysis indicates the relevance of strong-coupling effects~\cite{Profe2023s}. In contrast to a conventional Stoner instability, where a finite critical scale signals an instability already at weak to intermediate coupling, the absence of such a scale together with the slow (algebraic) growth of the eigenvalue implies that no instability is reached within the perturbative regime and that the flow is driven toward large effective interactions. This behavior points to a breakdown of the diagrammatic description and highlights the importance of local physics. To access this regime, we employ SBMF theory, which explicitly minimizes the free-energy landscape in the symmetry-broken phase (see \cref{ssec:sbmf}), and by that automatically includes the competition between different superpositions within the $E_2$ irrep subspace.

\subsubsection{Slave-Boson phase diagram}
The $U$--$t'$ phase diagram obtained from SBMF theory in \cref{fig:Phase_diagram_SBMF} is dominated by flat-band ferromagnetism at small $t'$, while a spin nematic phase (altermagnet, AM) appears for $-t' \gtrsim 0.1$, in agreement with the FM and $d$SPOM phases featured in the FRG phase diagram of~\cref{fig:Phase_diagram}.
The mean-field of the spin nematic phase has the form $(1,-1,0)$ in sublattice space, which indicates that two sublattices acquire the same magnitude of magnetization while the third remains paramagnetic (see $d$SPOM inset in~\cref{fig:Phase_diagram}). This corresponds to a $d$-wave AM spin texture on the kagome lattice~\cite{Peces2026a}.
Taking into account fluctuations around the correlated saddle point reveals that both the AM and FM states are unstable for $-t' \lesssim 0.2$.
The colored boundary line of this unstable regime in~\cref{fig:Phase_diagram_SBMF} marks the corresponding ordering vector $\bv{q}$ of the diverging transverse spin fluctuations. Charge and longitudinal spin susceptibilities remain at a moderate level.
Although $\bv{q}$ changes continuously along the phase boundary, it stays close to the $M$ point throughout the transition, largely consistent with the $M$-SDW found in the FRG analysis in this region. This behavior reflects a known tendency: Studies of the one-band Hubbard model have shown that SBMF tends to predict incommensurate magnetic ordering vectors for larger interactions, while FRG instabilities are typically locked to a nearby commensurate momentum~\cite{Klett2024i}.

A quantitative one-to-one correspondence between the FRG and the SBMF phase diagram is hampered by the absence of loop current orders in the latter method. This observation is expected because the Kotliar--Ruckenstein slave-boson ansatz employs local auxiliary fields and therefore naturally captures onsite spin and charge order parameters. In contrast, the (S)LCO phases identified in FRG correspond to bond orders arising from cross-channel feedback which is masked out in SBMF. While combinations of local SBMF degrees of freedom can influence their connecting bond correlations, the accessible parameter space remains significantly more restricted than in approaches that treat local and bond correlations on equal footing. For the conceptual extension of the SBMF ansatz to nonlocal order parameters see \cref{app:sbmf-bonds}. A comprehensive numerical analysis is beyond the scope of this work and will be provided elsewhere. Importantly, such extensions are further complicated by the Fierz ambiguity. As a result, the integration of nonlocal \emph{ansätze} requires a careful and systematic assessment of the relative weight and consistency of these channels in order to avoid biasing the mean-field solution toward a particular ordering tendency.

Upon introducing a nearest-neighbor density interaction $V$ at $t' = -0.3$ (right panel of \cref{fig:Phase_diagram_SBMF}), we do not observe an unstable regime---even though (S)LCO dominates the $U$--$V$ phase diagram recorded with FRG. This absence of an unstable regime further suggests that the saddle-point instability observed in the $U$--$t'$ plane should not be interpreted as a signature of bond-ordered tendencies.
In the large-$U/V$ and large-$V/U$ regimes, the SBMF calculations yield qualitative agreement with the FRG phase diagram:
Due to the absence of cross-channel projections, the AM state at large $U$ is solely driven by the spin exchange diagram in \cref{fig:diagrams}(a) and the phase boundary to the paramagnetic state is independent of NN interaction $V$.
At large $V$, we recover the intra-unit-cell charge-density wave (CDW) associated with the $d$POM instability in \cref{fig:Phase_diagram}.
Analyzing the evolution of the respective order parameters upon approaching the phase boundaries allows us to characterize the nature of the phase transitions.
We find that the onset of the AM order parameter occurs continuously, indicating second-order phase transitions as observed for, e.g., the altermagnetic transition on the Lieb lattice~\cite{Durrnagel2025a}.
\added{The phase transition from the parabolic metal to the (nematic) CDW state is first order, as expected from the allowed cubic term in the Ginzburg--Landau free energy ($A_1\subset E_2\otimes E_2\otimes E_2$).}
At large $U$ and $V$, the system transitions into a coexistence regime of AM and CDW order marked by abrupt changes in the order parameters indicative of first-order behavior.

\begin{figure*}
    \centering
    \includegraphics{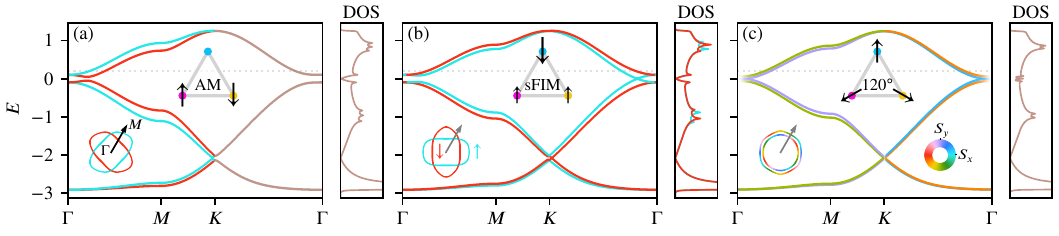}
    \caption{
    Spin-resolved pseudofermion band structures, Fermi surfaces, and density of states (DOS) for the altermagnetic (AM, a), the symmetry-compensated ferrimagnetic (sFIM, b), and the non-collinear $120^\circ$ (c) linear combinations of the $d$SPOM order parameter obtained from slave boson mean-field theory (SBMF) at $U=15$. We note that within SBMF, the three states are very close in energy, but the AM state (a) consistently minimizes the free energy $F$. In addition to band structure and DOS we show Fermi surfaces for slight particle doping away from $\nu=2/3$.
    While the collinear superpositions (a,b) feature Dirac crossings (linear DOS) at $E=0$, the $120^\circ$ order retains the initial parabolic band touching point (nonzero DOS).
    }
    \label{fig:E2_comparison}
\end{figure*}

\subsubsection{Free energy comparison between different $E_2$ states}
Even though AM and CDW as well as their leading scattering contributions in \cref{fig:diagrams}(a,b) are related via crossing symmetries $D \leftrightarrow C$ ($J \leftrightarrow V$), it is interesting to note that the respective orders realize different basis states of the two dimensional $E_2$ irrep associated with the transition. The CDW is given by the even basis vector [$(2,-1,-1)$ in sublattice space] with respect to the horizontal and vertical mirrors, while the AM realizes the odd one [$(0,1,-1)$]. This observation motivates to inspect the other magnetic states within the $E_2$ irrep and their energetic competition with the AM.
Since the magnetic order parameter is vector-valued, the relative orientation of the magnetic quantization axis of the eigenvectors provides an additional degree of freedom to minimize the free energy of the system and allows for both collinear and coplanar spin textures, e.g., $120^\circ$ order.
The overall spin orientation, though, remains arbitrary also in the ordered phase in the absence of spin-orbit coupling.

The system's ground state (within SBMF) is always given by the mirror odd basis vector of $E_2$ (i.e., the AM) and breaks $C_6$ symmetry.
Since the spin-$\uparrow$ sublattice is mapped on the spin-$\downarrow$ sublattice by the vertical and horizontal mirrors, this collinear state features vanishing net magnetization protected by $\{C_2 \parallel \sigma_{x/y} \}$ where the first operation acts in spin space and the second in real space.
This state can hence be classified as a $d_{xy}$-wave AM as also revealed by the spin-split band structure with symmetry protected nodal lines in \cref{fig:E2_comparison}(a)~\cite{Smejkal2022b, Smejkal2022e}.
The corresponding collinear mirror even eigenstate on the other hand is not compensated at all energies, see \cref{fig:E2_comparison}(b).
Even though the spin-split Fermi surface in \cref{fig:E2_comparison}(b) suggests labeling it as a $d_{x^2-y^2}$-wave AM, the nodal lines do not coincide with any symmetries of the underlying kagome lattice structure.
This is at odds with $d_{x^2-y^2}$-wave AMs in square lattice geometries~\cite{Roig2024m, Durrnagel2025a} and, in principle, allows for an occupation imbalance of spin-$\uparrow$ and spin-$\downarrow$ states in the quasiparticle band structure.
This is also apparent from the real space structure, where the different spin sublattices cannot be transformed into each other by any symmetry operation in $C_{6v}$.

\begin{figure}
    \centering
    \includegraphics[width=1\linewidth]{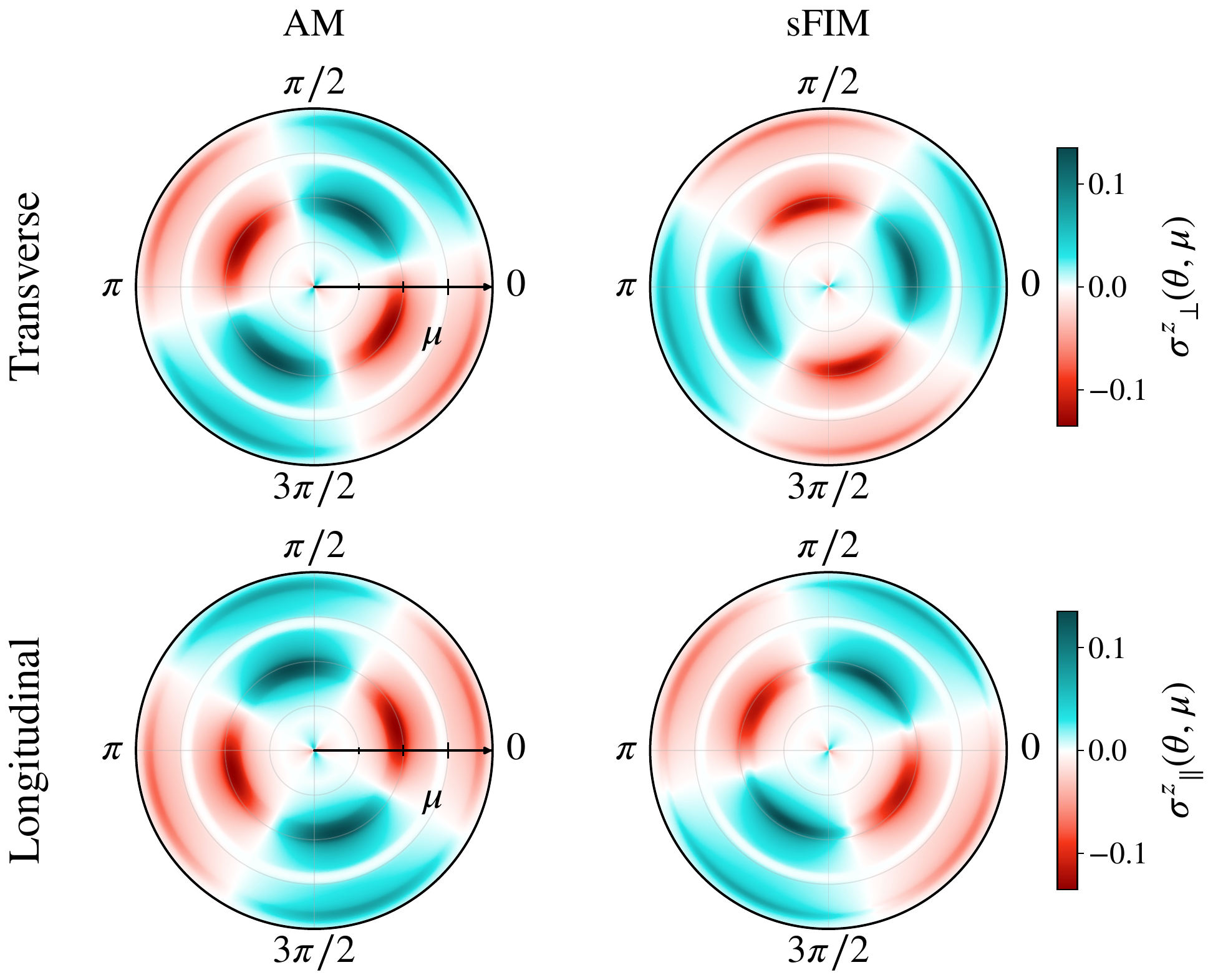}
    \caption{Angular spin-\(z\) transport response of the two compensated collinear orders as a function of chemical potential.
    The radial coordinate denotes \(\mu\), while the polar angle denotes the electric-field direction \(\theta\).
    The upper row shows the transverse response \(\sigma^{z}_{\perp}(\theta,\mu)\), and the lower row shows the longitudinal response \(\sigma^{z}_{\parallel}(\theta,\mu)\).
    Both magnetic configurations display a $d$-wave anisotropic spin-transport pattern, but with a relative \(\pi/4\) rotation.
    }
    \label{fig:Transport}
\end{figure}

Since the order parameter transforms in a non-trivial irrep, the magnetic expectation value sums up to zero within the unit cell regardless of its transformation behavior under elements of $C_{6v}$. In this sense, the mirror even $E_2$ basis state presents a symmetry compensated collinear ferrimagnet (sFIM).
In the vicinity of $\nu=2/3$ filling, this compensation even transcends to the spin-split band structure. The point-like Fermi surface ensures an equal occupation of both spin species by means of Luttinger's theorem~\cite{Luttinger1960f}.
Strict spin compensation on the quasi-particle level requires an additional filling constraint closely related to earlier proposals of fully compensated ferrimagnets~\cite{Wurmehl2006v,Mazin2022e,Liu2025t}. Even though perfect spin compensation at finite doping is lost due to the underlying $C_{6v}$ symmetry of the kagome lattice, the transport signatures of both the sFIM and AM states are mostly determined by the states close to the Fermi level.
Hence, transverse and longitudinal spin conductivities in \cref{fig:Transport} are governed by the spin splitting order parameter and show almost perfect $C_4$ symmetry between the AM and sFIM states---at arbitrary filling and even though $C_4$ is \emph{not} a lattice symmetry.
The only difference is the location of the spin nodes that follow either $d_{xy}$ or $d_{x^2-y^2}$ harmonics (for details see \cref{app:transport}).

Tracking the free energies of the AM and sFIM states (within SBMF) reveals a very close energetic competition.
This can be partially attributed to the small absolute scale of the order parameter even deep in the ordered phase (cf.~gap sizes in \cref{fig:E2_comparison} for $U=15$).
A comparison with the $120^\circ$ order, that relates to the chiral superposition of the two basis states, shows a clear separation in free energy.
This can be attributed to the different terms in the MF Hamiltonian that get minimized more efficiently depending on the order. While the $120^\circ$ order eases the geometric frustration of the NN spin exchange generated by the kinetic hopping $t$, AM and sFIM states are more effective in minimizing the bare interaction $U$.
This results in the following interesting phenomenology: Since the individual eigenstates within $E_2$ do not respect the underlying lattice symmetries, the transition from a paramagnetic to a magnetically ordered state is governed by energy minimization rather than symmetry.
This statement holds true for every order parameter in a multi-dimensional irrep and is at odds with the symmetry based classification of magnetic ground states based on spin space groups, where different members of a multi-dimensional irrep are categorized in distinct classes~\cite{Xiao2024s}. 

\section{Summary and Outlook}
We present the generalized kagome Hubbard model at $2/3$ filling as an ideal platform to study parabolic semimetals with nontrivial Bloch states.
Resummation of vertex corrections from the itinerant perspective via the functional renormalization group reveals a rich phase diagram largely at intermediate coupling.
The dominant phases are spin and charge current orders as well as spin and charge Pomeranchuk orders that are essentially driven by the presence of a quadratic band touching point in the absence of an extended Fermi surface.
A close competition between $U$-induced nematic spin fluctuations and nearest-neighbor Coulomb repulsion $V$ is revealed, resulting in favoring either LCO ($U$) or SLCO ($V$) at one loop level.
When approaching the flat band regime where $t'$ is small or when $V$ is large, higher order loop corrections reveal the inverse effect where additional inter-channel reinsertions dominate---\textit{a posteriori} justifying the usage of FRG as compared to simpler methods like RPA.
Close to the particle-hole symmetric regime $t'=-t/3$ and at larger $U$, we find an enhancement of nematic spin exchange that favors spin Pomeranchuk order.
The inter-channel diagrammatic correction entailed by the FRG treatment render the divergence algebraic, which stands at odds with Stoner-like instabilities typically encountered in FRG (and, of course, RPA).
This indication necessitates a strong coupling method to access ordering tendencies at nonzero temperatures, which we provide in the form of auxiliary particles (slave boson mean-fields, SBMF).
The slave boson analysis corroborates the result of both spin and charge Pomeranchuk order and demonstrates how the two states are traced to strong coupling ($U, V\gg t,t'$).
SBMF naturally minimizes the free energy inside the symmetry-broken regime, allowing us to differentiate between different spin configurations included in the two dimensional irrep describing (spin) Pomeranchuk order.
We show that the ultimately favored state is a collinear altermagnet, which is in close energetic competition with the related symmetry-compensated ferrimagnet. In contrast, the $120^\circ$ non-collinear altermagnet is suppressed.
We note that the typical spin space group classification of magnetic orders hides the shared origin (i.e., same irrep of parent group) of the AM and sFIM states.

We finally would like to comment on the ubiquitous appearance of (S)LCO in the kagome Hubbard model at $\nu=2/3$ filling, which stands in stark contrast to Van Hove filling. There, LCO states have been proposed as subsidiary order to a charge bond order parent phase~\cite{guo2024correlated}. They nevertheless remain notoriously hard to stabilize as leading instability in any unbiased many-body treatment and instead necessitate explicit---potentially uncontrolled---diagrammatic cross-insertions~\cite{Tazai2023c}. Only recently, a Haldane-like LCO has been found to be the leading instability in a spinless model with long-range interactions~\cite{Zhan2026l}.
The parabolic semimetallic nature in conjunction with nontrivial quantum geometry places the $\nu=2/3$ filled kagome Hubbard model in a regime where one-loop diagrams support both charge and spin loop currents, with close competition between the two upon inclusion of higher order diagrams as evidenced by our FRG analysis.


\begin{acknowledgments}
We thank J.~Ingham for collaborations on related topics \added{and valuable feedback on the manuscript,} and J.~Seufert for fruitful discussions. 
This research was funded by the Deutsche Forschungsgemeinschaft (DFG, German
Research Foundation) -- Project-ID 258499086 -- SFB 1170; through the
Würzburg-Dresden Cluster of Excellence on Complexity, Topology, and Dynamics in
Quantum Materials (ctd.qmat) -- Project-ID 390858490 -- EXC 2147; and through
the Research Unit QUAST -- Project-ID 449872909 -- FOR 5249.
M.D. is grateful for support from a Ph.D. scholarship of the Studienstiftung des deutschen Volkes.
\end{acknowledgments}


\appendix
\crefalias{section}{appendix}

\section{FRG details}
\label{app:frg-details}
The phase diagrams (\cref{fig:Phase_diagram}) are sampled with an initial resolution of $6\times8$ ($t'\times U$) and $5\times8$ ($V\times U$) points and refinement (in three factor two recursion steps on a rectangular grid) along the phase boundaries. This approach allows to sample less than $1/4$ of all points in phase space. The critical scale in \cref{fig:Phase_diagram} is interpolated logarithmically for points not sampled by the recursive phase boundary algorithm.

Each of the FRG runs is carried out using the open source {divERGe} library~\cite{10.21468/SciPostPhysCodeb.26, 10.21468/SciPostPhysCodeb.26-r0.5} with the TUFRG backend. The bosonic (fermionic) momenta are resolved on a regular $48\times48$ ($2304\times2304$) grid in the primitive zone. We truncate the TU expansion at a distance of two lattice vectors. The flow is terminated when a vertex eigenvalue in the physical channels (charge, magnetic, superconducting; not $P$, $C$, $D$) reaches a value larger than $30$ (in units of $t$). We use {divERGe}'s internal Euler forward integrator to solve the flow equation and set its parameters to:
\begin{itemize}[itemsep=-.3em]
    \newcommand\eulerParam[2]{\item\texttt{#1 = #2;}}
    \eulerParam{eu.Lambda}{100}
    \eulerParam{eu.dLambda\_fac}{0.1}
    \eulerParam{eu.dLambda\_fac\_scale}{30.0}
    \eulerParam{eu.consider\_maxvert\_lambda}{5.0}
    \eulerParam{eu.lambda\_min}{1.e-4}
\end{itemize}

The instabilities are determined by diagonalizing the \emph{physical channels}, i.e., charge ($2V^D-V^C$), magnetic ($-V^C$), and superconducting ($V^P$) at every transfer momentum $\bv q$, finding the overall leading eigenvalue, and inspecting the corresponding eigenvector(s) for their form factor content.

\section{SBMF details}
\label{app:sbmf-details}
The SBMF phase diagrams in \cref{fig:Phase_diagram_SBMF} as well as the static susceptibilities are calculated at $T=0.01$ and the Brillouin zone is sampled with $512\times512$ regularly spaced points. For the free energy comparison of the different $E_2$ states this is increased to $1024 \times 1024$. We refer the reader to Ref.~\cite{Riegler2020s} for a comprehensive overview over the SBMF method.

\section{Transport signatures}
\label{app:transport}
\subsubsection{Spin-resolved Kubo formula and Drude limit}

For the collinear magnetic states considered here, the spin quantization axis is
fixed by the ordered moment. The natural spin projection is therefore the fixed
$z$-spin projection, $P_z^\pm=(1\pm\sigma_z)/2$, where $+$ and $-$
denote the two spin sectors. The velocity operator is
$\hat v_i(\bv k)=\hbar^{-1}\partial_{k_i}H(\bv k)$, with $i=x,y$.
Formally, the projected current in each spin sector is
$\hat J_i^\pm(\bv k)=\{\hat v_i(\bv k),P_z^\pm\}/2$. Since the
collinear Hamiltonian is block diagonal in $\sigma_z$, the velocity operator
commutes with $P_z^\pm$. Hence
$\hat J_i^\pm(\bv k)=\hat v_i(\bv k)P_z^\pm$, and the response can be
evaluated independently in the two spin blocks. For each spin sector, the multi-band Kubo response \cite{Kubo1957k,Greenwood1958g,Huhtinen2023c,Calderin2017k} is computed from
\begin{equation}
\begin{aligned}
\sigma_{ij}^{s}(\omega)
&\propto
\sum_{bb'\bv k}
\frac{
f\!\left(\epsilon_{b,s}(\bv k)\right)
-
f\!\left(\epsilon_{b',s}(\bv k)\right)
}{
\epsilon_{b,s}(\bv k)-\epsilon_{b',s}(\bv k)
}
\\
&\hspace{1.0cm}\times
\frac{
v_{i,s}^{bb'}(\bv k)
v_{j,s}^{b'b}(\bv k)
}{
\hbar\omega
+
\epsilon_{b,s}(\bv k)
-
\epsilon_{b',s}(\bv k)
+i\eta
}.
\end{aligned}
\label{eq:spin_block_kubo}
\end{equation}
Here $f(\epsilon)$ is the
Fermi--Dirac distribution and $v_{i,s}^{bb'}(\bv k)
=
\bra{u_{b,s}(\bv k)}
\hat v_i(\bv k)
\ket{u_{b',s}(\bv k)}$
are velocity matrix elements in the band basis. The full expression in
\cref{eq:spin_block_kubo} was evaluated in the low-temperature ($T\rightarrow 0$) static limit.
For the collinear states considered here, this calculation gives no qualitative
change relative to the intraband result, so the transport patterns shown below
are governed by the Drude contribution. The
relaxation time $\tau$ phenomenologically accounts for elastic scattering and
replaces the infinitesimal broadening in the dc intraband limit. The $b=b'$ part of
\cref{eq:spin_block_kubo} gives the intraband contribution and taking the static low-temperature limit yields the usual spin-block Drude expression
\cite{Culcer2004s,Shi2006p,Zhang2008t,Calixto2026r},
\begin{equation}
\begin{aligned}
\sigma_{ij}^{s}(\mu)
&=
e^2\tau
\sum_n
\int_{\mathrm{BZ}}\!\frac{d^2k}{(2\pi)^2}\,
v_{i,s}^{(n)}(\bv k)
v_{j,s}^{(n)}(\bv k)
\\
&\hspace{1.0cm}\times
\delta\!\left(E_{n,s}(\bv k)-\mu\right),
\qquad s=\uparrow,\downarrow .
\end{aligned}
\label{eq:spin_block_drude}
\end{equation}
with
\begin{equation}
v_{i,s}^{(n)}(\bv k)
=
\frac{1}{\hbar}\partial_{k_i}E_{n,s}(\bv k).
\end{equation}
The spin-$z$ conductivity tensor is defined as
\begin{equation}
\boldsymbol{\sigma}^{z}
=
\boldsymbol{\sigma}^{\uparrow}
-
\boldsymbol{\sigma}^{\downarrow}.
\end{equation}

\subsubsection{Angular transport response}

To characterize the angular response, we apply an electric field along
\(\hat{\bv n}(\theta)=(\cos\theta,\sin\theta)\). The transverse response is
measured along the perpendicular direction
\(\hat{\bv m}(\theta)=(-\sin\theta,\cos\theta)\), while the longitudinal
response is measured along \(\hat{\bv n}(\theta)\). Thus
\(\sigma^z_\perp(\theta)=
\hat{\bv m}(\theta)\cdot\boldsymbol{\sigma}^z\hat{\bv n}(\theta)\)
and
\(\sigma^z_\parallel(\theta)=
\hat{\bv n}(\theta)\cdot\boldsymbol{\sigma}^z\hat{\bv n}(\theta)\). For the symmetric Drude tensor obtained in the present relaxation-time
calculation, \(\sigma^z_{xy}=\sigma^z_{yx}\), these angular responses become
\begin{equation}
\begin{aligned}
\sigma^z_\perp(\theta)
&=
\sigma^z_{xy}\cos 2\theta
+
\frac{\sigma^z_{yy}-\sigma^z_{xx}}{2}
\sin 2\theta ,
\\
\sigma^z_\parallel(\theta)
&=
\frac{\sigma^z_{xx}+\sigma^z_{yy}}{2}
+
\frac{\sigma^z_{xx}-\sigma^z_{yy}}{2}
\cos 2\theta
+
\sigma^z_{xy}\sin 2\theta .
\end{aligned}
\label{eq:angular_transport_symmetric}
\end{equation}
The anisotropic part is therefore controlled by the two tensor components
\((\sigma^z_{xx}-\sigma^z_{yy})/2\) and \(\sigma^z_{xy}\), which form an
\(E_2\)-type response. The sFIM and AM states realize two nearly orthogonal
orientations of this \(E_2\) representation. Consequently, their angular spin-transport
patterns are shifted relative to one another: when one state is dominated by the
\(\sin2\theta\) harmonic and the other by the \(\cos2\theta\) harmonic, the
transverse response is rotated by \(\pi/4\). The longitudinal response exhibits
the same relative shift, with the sine and cosine harmonics interchanged. This
relative phase shift provides a direct transport signature distinguishing the
sFIM and AM states in ~\cref{fig:Transport}.

\section{Bond orders in SBMF}
\label{app:sbmf-bonds}

In the main text, the SBMF treatment is based on a
local Kotliar--Ruckenstein saddle point. Such an ansatz naturally captures
onsite charge and spin order, but does not generate an independent bond-order
field. To make this limitation explicit, we briefly formulate the corresponding
bond-channel extension. In the spin-rotationally invariant Kotliar--Ruckenstein representation, the
physical electron is written as
\begin{equation}
    c^\dagger_{i\sigma}
    =
    \sum_{\varsigma}
    z^\dagger_{i,\sigma\varsigma}
    f^\dagger_{i\varsigma},
    \qquad
    c_{i\sigma}
    =
    \sum_{\varsigma}
    z_{i,\sigma\varsigma}
    f_{i\varsigma}.
    \label{eq:electron_sbmf_app}
\end{equation}
Here \(f_{i\sigma}\) is a pseudofermion and \(\hat z_i\) is the local
quasiparticle-renormalization matrix. The latter is constructed from the local
slave-boson amplitudes \(e_i\), \(d_i\), \(p_{0,i}\), and \(p_{\mu,i}\)
with \(\mu=x,y,z\). The fields \(e_i\) and \(d_i\) describe empty and doubly
occupied configurations, while \(p_{0,i}\) and \(p_{\mu,i}\) describe the
spin-singlet and spin-vector components of the singly occupied sector. These
auxiliary fields enlarge the local Hilbert space and must therefore be
constrained back to the physical subspace. At saddle-point level, the relevant
constraints are
\begin{align}
    1
    &=
    e_i^2+d_i^2+p_{0,i}^2+\sum_{\mu=x,y,z}p_{\mu,i}^2,
    \label{eq:sb_completeness_app}
    \\
    n_i^f
    &=
    p_{0,i}^2+\sum_{\mu=x,y,z}p_{\mu,i}^2+2d_i^2,
    \label{eq:sb_density_constraint_app}
\end{align}
together with the spin constraints
\begin{equation}
    \sum_{\sigma\sigma'}
    f^\dagger_{i\sigma'}
    (\tau_\mu)_{\sigma'\sigma}
    f_{i\sigma}
    =
    \mathcal P_{i,\mu},
    \qquad
    \mu=x,y,z,
    \label{eq:sb_spin_constraint_app}
\end{equation}
where \(\mathcal P_{i,\mu}\) denotes the corresponding bosonic spin expression.
The constraints are enforced by Lagrange multipliers contained in
\(H_{\mathrm{constr}}\). The local physical density is therefore
\begin{equation}
    n_i^{\mathrm{SB}}
    =
    p_{0,i}^2+\sum_{\mu=x,y,z}p_{\mu,i}^2+2d_i^2 .
    \label{eq:sb_density_app}
\end{equation}

\subsubsection{Limitations of local ansatz for bond order}
It is instructive to consider whether bond/current-like structures can already be generated within the local KR ansatz by using purely vectorial slave-boson
fields. The spin-rotationally invariant \(z\)-matrix can be written as
\begin{equation}
    z_i
    =
    z_i^{0}\tau_0
    +
    z_i^{p}\,
    \hat{\bv p}_i\cdot\boldsymbol{\tau},
    \qquad
    \hat{\bv p}_i
    =
    \frac{\bv p_i}{|\bv p_i|},
    \label{eq:z_decomp_pure_p}
\end{equation}
where
\begin{equation}
    z_i^{0}
    =
    \frac{z_{i,+}+z_{i,-}}{2},
    \qquad
    z_i^{p}
    =
    \frac{z_{i,+}-z_{i,-}}{2}.
\end{equation}
Here \(z_{i,\pm}\) are the two quasiparticle weights associated with the local
spin quantization axis set by \(\bv p_i\). Explicitly,
\begin{multline}
    z_{i,\pm}
    =
    \big(p_{0,i}(e_i+d_i)
    \pm
    |\bv p_i|(e_i-d_i)\big) \\
    {} \times \bigg[
    2 \sqrt{
    1-d_i^2-(p_{0,i}\pm|\bv p_i|)^2/2
    } \\
    \sqrt{
    1-e_i^2-(p_{0,i}\mp|\bv p_i|)^2/2
    }
    \bigg]^{-1}
    \,.
    \label{eq:zpm_app}
\end{multline}
For a pure \(p\)-vector configuration,
\begin{equation}
    p_{0,i}=0,
    \qquad
    \bv p_i\neq0,
    \label{eq:pure_p_condition}
\end{equation}
the two weights satisfy \(z_{i,+}=-z_{i,-}\). Hence the scalar part vanishes,
\(z_i^{0}=0\), and the local renormalization matrix becomes purely vectorial,
\begin{equation}
\begin{aligned}
    \hat z_i
    &{} =
    z_i^{p}\,
    \hat{\bv p}_i\cdot\boldsymbol{\tau},
    \\
    z_i^{p}
    &{} =
    \frac{
    |\bv p_i|(e_i-d_i)
    }{
    2
    \sqrt{1-d_i^2-|\bv p_i|^2/2}
    \sqrt{1-e_i^2-|\bv p_i|^2/2}
    } .
\end{aligned}
    \label{eq:z_pure_p}
\end{equation}
At the same time, the local spin moment vanishes because
\begin{equation}
    \bv S_i^{\mathrm{SB}}
    =
    p_{0,i}
    \begin{pmatrix}
    p_{x,i}\\
    -p_{y,i}\\
    p_{z,i}
    \end{pmatrix}
    =
    0 .
    \label{eq:pure_p_spin_zero}
\end{equation}
Thus a pure \(p\)-vector saddle point is locally nonmagnetic, although the
quasiparticle weight matrix is spin dependent.

The effect on a hopping process follows from
\begin{equation}
    \hat T_{ij}^{\mathrm{SB}}
    =
    \hat z_i^\dagger
    t_{ij}\tau_0
    \hat z_j .
\end{equation}
Using \cref{eq:z_pure_p}, one obtains
\begin{equation}
\begin{aligned}
    \hat T_{ij}^{\mathrm{SB}}
    &{}=
    t_{ij} z_i^{p} z_j^{p}
    \left(
    \hat{\bv p}_i\cdot\boldsymbol{\tau}
    \right)
    \left(
    \hat{\bv p}_j\cdot\boldsymbol{\tau}
    \right)
    \\
    &{}=
    t_{ij} z_i^{p} z_j^{p}
    \left[
    \left(
    \hat{\bv p}_i\cdot\hat{\bv p}_j
    \right)\tau_0
    +
    i
    \left(
    \hat{\bv p}_i\times\hat{\bv p}_j
    \right)\cdot\boldsymbol{\tau}
    \right].
\end{aligned}
\label{eq:pure_p_hopping}
\end{equation}
Therefore, noncollinear pure \(p\)-vectors generate an effective spin-dependent
bond hopping. For example, if
\(\hat{\bv p}_i=\hat{\bv x}\) and
\(\hat{\bv p}_j=\hat{\bv y}\), then
\begin{equation}
    \hat T_{ij}^{\mathrm{SB}}
    =
    i\,t_{ij}z_i^{p}z_j^{p}\tau_z ,
    \label{eq:pure_p_ixy}
\end{equation}
whereas reversing the bond gives the Hermitian conjugate. Such a structure
resembles the spin-dependent imaginary hopping expected for a SLCO
pattern.

However, this construction is not an independent bond order. The bond matrix in
\cref{eq:pure_p_hopping} is fully determined by local quantities,
\(z_i^p\), \(z_j^p\), \(\hat{\bv p}_i\), and \(\hat{\bv p}_j\), which
are themselves constrained by
\begin{equation}
    e_i^2+d_i^2+|\bv p_i|^2=1,
    \qquad
    n_i^{\mathrm{SB}}
    =
    |\bv p_i|^2+2d_i^2
    \quad
    (p_{0,i}=0).
    \label{eq:pure_p_constraints}
\end{equation}
Consequently, the amplitude and spin structure of the bond hopping cannot be
varied independently from the onsite slave-boson configuration. In our local
SBMF saddle-point analysis, such pure \(p\)-vector configurations are therefore
not stabilized as separate phases. They can mimic certain spin-dependent
hopping structures, but they do not constitute a general variational
description of the LCO or SLCO phases. 

\subsubsection{Bond order ansatz in conjunction with SBMF}

To include a possible bond channel, we split the nearest-neighbor interaction
into a Fock-decoupled bond part and a residual density-density part,
\begin{multline}
    H_V
    \rightarrow
    H_V^{(x)}
    =
    -xV
    \sum_{\langle ij\rangle}
    \sum_{\sigma\sigma'}
    c^\dagger_{i\sigma}
    c_{j\sigma'}
    c^\dagger_{j\sigma'}
    c_{i\sigma}
    \\ {} + (1-x)V
    \sum_{\langle ij\rangle}
    n_i^{\mathrm{SB}}n_j^{\mathrm{SB}} .
    \label{eq:HV_split_app}
\end{multline}
Here \(x\in[0,1]\) parametrizes how much of the nearest-neighbor interaction is treated in the Fock channel, i.e., acting on bonds~\footnote{The fact that $x\in[0,1]$ can be chosen freely reflects the Fierz ambiguity of several channels to decompose the interaction into.}. As this choice is a priori arbitrary, it introduces a bias between bond and on-site ordering tendencies, so one should analyze how the results depend on \(x\) and assess the stability of the resulting phases, for instance by considering Gaussian fluctuations.
The local SBMF calculation used in the main
text corresponds to \(x=0\). The Fock term is decoupled by introducing the spin-resolved bond expectation
value
\begin{equation}
    \gamma_{ij}^{\sigma\sigma'}
    =
    xV
    \braket{
    c^\dagger_{j\sigma'}c_{i\sigma}
    } ,
    \qquad
    \gamma_{ji}^{\sigma'\sigma}
    =
    \big(\gamma_{ij}^{\sigma\sigma'}\big)^* .
    \label{eq:gamma_def_app}
\end{equation}
Neglecting quadratic fluctuations around this expectation value gives
\begin{multline}
    H_{V,\mathrm{bond}}^{\mathrm{MF}}
    =
    -
    \sum_{\langle ij\rangle}
    \sum_{\sigma\sigma'}
    \gamma_{ij}^{\sigma\sigma'}
    c^\dagger_{i\sigma}c_{j\sigma'}
    +\mathrm{h.c.}
    \\
    {}+
    \frac{1}{xV}
    \sum_{\langle ij\rangle}
    \sum_{\sigma\sigma'}
    \gamma_{ij}^{\sigma\sigma'}
    \gamma_{ji}^{\sigma'\sigma}.
    \label{eq:bond_mf_app}
\end{multline}
Equivalently, using spinor notation
\(c_i=(c_{i\uparrow},c_{i\downarrow})^T\) and the \(2\times2\) matrix
\((\hat\gamma_{ij})_{\sigma\sigma'}=\gamma_{ij}^{\sigma\sigma'}\), this reads
\begin{equation}
    H_{V,\mathrm{bond}}^{\mathrm{MF}}
    =
    -
    \sum_{\langle ij\rangle}
    c_i^\dagger \hat\gamma_{ij} c_j
    +\mathrm{h.c.}
    +
    \frac{1}{xV}
    \sum_{\langle ij\rangle}
    \mathrm{Tr}\,
    \hat\gamma_{ij}\hat\gamma_{ji}.
    \label{eq:bond_mf_matrix_app}
\end{equation}

Substituting the slave-boson representation
\cref{eq:electron_sbmf_app} into \cref{eq:bond_mf_matrix_app} gives
\begin{equation}
    H_{V,\mathrm{bond}}^{\mathrm{MF}}
    =
    -
    \sum_{\langle ij\rangle}
    f_i^\dagger
    \hat z_i^\dagger
    \hat\gamma_{ij}
    \hat z_j
    f_j
    +\mathrm{h.c.}
    +
    \frac{1}{xV}
    \sum_{\langle ij\rangle}
    \mathrm{Tr}\,
    \hat\gamma_{ij}\hat\gamma_{ji}.
    \label{eq:bond_mf_sbmf_app}
\end{equation}
Thus the bond field enters the pseudofermion Hamiltonian as an additional
interaction-generated hopping matrix. The kinetic hopping is renormalized in
the same way,
\begin{equation}
    H_t^{\mathrm{SBMF}}
    =
    -
    \sum_{\langle ij\rangle}
    f_i^\dagger
    \hat z_i^\dagger
    t_{ij}\tau_0
    \hat z_j
    f_j
    +\mathrm{h.c.},
    \label{eq:sb_hopping_app}
\end{equation}
so that the combined hopping sector becomes
\begin{multline}
    H_{t+\gamma}^{\mathrm{SBMF}}
    =
    -
    \sum_{\langle ij\rangle}
    f_i^\dagger
    \hat z_i^\dagger
    \left(
    t_{ij}\tau_0+\hat\gamma_{ij}
    \right)
    \hat z_j
    f_j
    +\mathrm{h.c.}
    \\
    {}+
    \frac{1}{xV}
    \sum_{\langle ij\rangle}
    \mathrm{Tr}\,
    \hat\gamma_{ij}\hat\gamma_{ji}.
    \label{eq:combined_bond_sbmf_app}
\end{multline}

The full saddle-point Hamiltonian with this bond-channel extension can be
written schematically as
\begin{multline}
    H_{\mathrm{SBMF}}^{(x)}
    =
    H_{t+\gamma}^{\mathrm{SBMF}}
    +
    U\sum_i d_i^2
    -
    \mu\sum_i n_i^f
    \\
    {}+
    (1-x)V
    \sum_{\langle ij\rangle}
    n_i^{\mathrm{SB}}n_j^{\mathrm{SB}}
    +
    H_{\mathrm{constr}} .
    \label{eq:full_sbmf_bond_app}
\end{multline}
The constraint contribution contains the local completeness, density, and spin
constraints,
\begingroup
\allowdisplaybreaks[1]
\begin{multline}
    H_{\mathrm{constr}}
    =
    \sum_i
    \alpha_i
    \left[
    1-e_i^2-d_i^2-p_{0,i}^2
    -\sum_{\mu}p_{\mu,i}^2
    \right]
    \\ 
    {}+
    \sum_i
    \beta_{0,i}
    \left[
    n_i^f
    -
    p_{0,i}^2
    -
    \sum_{\mu}p_{\mu,i}^2
    -
    2d_i^2
    \right]
    \\
    {}+
    \sum_i
    \sum_{\mu=x,y,z}
    \beta_{\mu,i}
    \left[
    \sum_{\sigma\sigma'}
    f^\dagger_{i\sigma'}
    (\tau_\mu)_{\sigma'\sigma}
    f_{i\sigma}
    -
    \mathcal P_{i,\mu}
    \right].
    \label{eq:Hconstr_app}
\end{multline}
\endgroup

Equation~\eqref{eq:combined_bond_sbmf_app} shows that a genuine bond order
requires an additional variational field \(\hat\gamma_{ij}\). In the local KR
ansatz used in the main text, no such independent bond field is introduced.
The only bond dependence then arises indirectly through products of local
renormalization matrices, \(\hat z_i^\dagger t_{ij}\hat z_j\). For a uniform
paramagnetic saddle point, \(\hat z_i=z_0\tau_0\), this merely rescales the
bare hopping,
\begin{equation}
    \hat z_i^\dagger t_{ij}\tau_0 \hat z_j
    =
    z_0^2 t_{ij}\tau_0 .
\end{equation}
More general local spin-dependent saddle points can make
\(\hat z_i^\dagger t_{ij}\hat z_j\) spin dependent, but this structure remains
fully determined by onsite bosonic amplitudes. It therefore cannot represent an
arbitrary bond-current or spin-bond-current pattern, independently of the local
charge and spin configuration. This is why the local SBMF calculation is well
suited to the onsite \(E_2\) charge and spin Pomeranchuk orders, but does not
provide a complete variational description of the LCO and SLCO phases found in
FRG.

\bibliography{literature.bib}

\end{document}